\documentclass[preprintnumbers, twocolumn, showpacs]{revtex4}
\usepackage{graphicx}
\usepackage{fancyhdr}
\usepackage{pslatex}
\usepackage{amsfonts}
\usepackage[utf8]{inputenc}
\usepackage[mathcal]{eucal}
\usepackage{latexsym}
\usepackage{enumerate}

\begin{document}

\title{Ondas sonoras estacionárias em um tubo: análise de problemas \& sugestões \newline
\small{\textit{ (Standing Sound Waves in a tube: Approach analysis \& sugestions )}   }  }

\author{L. P. Vieira$^{\dag}$, D. F. Amaral $^{\ddag}$ and V. O. M. Lara$^{\star}$}
\affiliation{
$^{\dag}$ Instituto de F\'isica - Universidade Federal do Rio de Janeiro, Rio de Janeiro - Rio de Janeiro Brasil\\
$^{\ddag}$ Consórcio de Ensino à distância do Rio de Janeiro (CEDERJ), pólo São Gonçalo -  Rio de Janeiro Brasil  \\
$^{\star}$ Instituto de F\'isica - Universidade Federal Fluminense, Niter\'oi - Rio de Janeiro Brasil \\
}

\pacs{ 
}

\date{\today}

\begin{abstract}

No presente trabalho temos como objetivo apresentar alguns questionamentos com respeito à abordagem utilizada em alguns livros didáticos de nível médio sobre o tema de ondas sonoras estacionárias em tubos. Além de classificar os livros didáticos dentro de um conjunto de critérios estabelecidos, apresentamos também algumas sugestões para uma discussão mais aprofundada deste tema. Sugerimos o uso de gifs e animações e a utilização de dois experimentos simples, que permitem a visualização dos perfis de variação de pressão e deslocamento de ar para os modos harmônicos de vibração.
\par
\textbf{Palavras-chave}: Ondas sonoras estacionárias, Análise de livros didáticos, \textit{Tablets} e \textit{smartphones}, Uso de Tecnologias no Ensino de Ciências.

In this paper we attempt to present some questions with respect to the approach used in some brazilian mid-level textbooks on the topic of stationary sound waves in tubes. In addition to ranking the textbooks within a set of criteria, we also present some suggestions for further discussions of this topic. We suggest the use of gifs and animations and the use of two experiments that allow you to view the profiles of variation of pressure and air displacement for the harmonic modes of vibration.

\end{abstract}







\maketitle

\vskip \baselineskip

\section{Introdu\c{c}\~ao}

Diversos estudos mostram que a utilização de experimentos em sala de aula são de grande importância na aprendizagem dos conteúdos apresentados nas áreas das Ciências da Natureza \cite{bybee, krasilchik, longhini}. Infelizmente a inexistência de um espaço físico e/ou a falta de infraestrutura prejudicam a prática destas atividades experimentais. Este cenário é bastante comum nas escolas públicas e em uma boa parcela das escolas particulares \cite{borges}.

Alguns trabalhos, tais como \cite{previous_mag, accelerometer, artigo_gota}, já demonstraram a gama de aplicativos que tais dispositivos apresentam e que podem ser explorados para fins didáticos. Em particular na área da Física, muitas leis podem ser confirmadas e visualizadas, o que torna alguns conceitos físicos mais concretos e palatáveis.

Tendo isto em mente, apresentamos neste trabalho dois experimentos que podem ser utilizados para se discutir ondas sonoras estacionárias em um tubo semi aberto. Embora tenhamos nos restringido ao caso de tubos semiabertos por brevidade, nada impede o uso dos experimentos apresentados aqui em um tubo aberto, enquanto que o caso de um tubo fechado exigiria uma sofisticação um pouco maior. Além disto, realizamos também uma avaliação da maneira com que este assunto é discutido em diversos livros de Física de nível médio. Conforme veremos, o tratamento realizado em boa parte dos livros peca em diversos aspectos, tais como a falta de clareza à respeito das quantidades físicas medidas, a natureza das ondas sonoras no ar (boa parte dos livros apresenta perfis relacionados à ondas transversais sem maiores preocupações), que, conforme sabemos, são longitudinais, e o comportamento oscilatório deste tipo de onda.

\section{Discussão Teórica}

O estudo de ondas sonoras em um tubo é assunto delicado devido à uma série de fatores. O problema surge quando desejamos representar ondas sonoras estacionárias em um livro \cite{site_d_russell}. Conforme sabemos, as ondas sonoras no ar são longitudinais. Entretanto, ao se discutir o caso de ondas sonoras estacionárias em um tubo, praticamente todos os livros apresentam imagens estáticas cujo formato é basicamente o de uma onda em uma corda esticada (como a corda de um violão), sendo associada portanto à uma onda transversal. Isto pode gerar uma série de dúvidas conceituais, fazendo com que o leitor incauto acredite que as imagens são a onda sonora em si, e não a representação esquemática da variação de uma grandeza física específica associada à esta onda sonora.

Outro problema deve-se ao fato de as ondas sonoras estacionárias em um tubo apresentarem oscilações temporais. As imagens ilustram apenas a amplitude de uma grandeza física em específico, no caso o deslocamento de ar. Conforme veremos posteriormente, podemos visualizar a oscilação temporal tanto para o deslocamento de ar (utilizando bolinhas de isopor) quanto para a variação de pressão (que está relacionada à intensidade da onda sonora) utilizando-se uma montagem experimental simples, ou fazendo uso de gifs e animações \cite{site_d_russell}.

Outro ponto bastante importante, e que não recebe a devida atenção em nenhum dos livros relacionados neste trabalho relaciona-se à uma aproximação fundamental que deve ser feita. Todo o tratamento feito à esse respeito deve supor que as ondas sonoras emitidas pela fonte sonora são ondas planas. Se a fonte sonora puder ser considerada uma fonte pontual, por exemplo, a discussão feita só será válida no regime em que estivermos suficientemente afastados da fonte, de modo que possamos considerar que as ondas sonoras no interior do tubo são essencialmente planas. Se uma fonte sonora pontual for posta na extremidade aberta de um tubo, por exemplo, o fenômeno físico seria consideravelmente mais complicado, uma vez que teríamos sucessivas reflexões nas paredes do tubo, dado o caráter esférico da onda emitida pela fonte.

\section{Avaliação dos livros}
\label{ava_livro}

A fim de classificar adequadamente os livros, elaboramos cinco critérios, enumerando-os de $1$ a $5$:

\begin{enumerate}
\item Quando apresenta as ondas sonoras estacionárias em um tubo, o livro salienta a natureza longitudinal deste tipo de onda?
\item O livro deixa claro qual é a quantidade física que está sendo representada (deslocamento de ar, ou mesmo "vibração da coluna de ar")?
\item O livro levanta a possibilidade de se medir outras quantidades (variação de pressão, ou intensidade da onda sonora)?
\item O livro discute o caráter temporal oscilatório da onda sonora estacionária, deixando claro que a imagem representa a amplitude do deslocamento de ar?
\item Apresenta a condição fundamental de que as ondas sonoras no tubo devem ser ondas planas?
\end{enumerate}

Tendo em vista os critérios listados àcima, elaboramos a Tabela I.

\begin{center}
    \begin{tabular}{ | l | l | l | l | l | l |p{5cm} |}
    \hline
     \textbf{Livros} & \textbf{1} & \textbf{2} & \textbf{3} & \textbf{4} & \textbf{5}\\ \hline
    Máximo \& Alvarenga \cite{maximo_alvarenga}& \checkmark & \checkmark & $\times$ & $\times$ & $\times$\\ \hline
    Guimarães \& Fonte Boa \cite{guimaraes_boa}& \checkmark & \checkmark & $\times$ & $\times$ & $\times$\\ \hline
    Ramalho et al \cite{ramalho}& $\times$ & \checkmark & $\times$ & $\times$ & $\times$\\ \hline
    Gaspar \cite{gaspar} & $\times$ & $\times$ & $\times$ & $\times$ & $\times$\\ \hline
    Helou et al \cite{helou}& \checkmark & \checkmark & \checkmark & $\times$ & $\times$ \\ \hline
    Hewitt * \cite{hewitt}& $\times$ & $\times$ & $\times$ & $\times$ & $\times$\\ \hline  
    Gref * \cite{gref} & $\times$ & $\times$ & $\times$ & $\times$ & $\times$\\ \hline
    \end{tabular}
    \label{table1}
\end{center}
\noindent
\textbf{Tabela I}: {\footnotesize Tabela que classifica os livros selecionados de acordo com os critérios apontados no texto. Os livros cujos autores aparecem com * indicam que o livro em questão não discute especificamente o caso de ondas sonoras estacionárias em um tubo.} 
\vspace{0.2 cm}		

De todos os livros avaliados nesta pesquisa bibliográfica, apenas a coleção de Helou et al diferencia o deslocamento de ar e a variação de pressão, apontando inclusive a defasagem de $90^{\circ}$ existente entre ambos os perfis. 

Conforme pode-se ver na Tabela I, boa parte dos livros salienta a natureza longitudinal das ondas sonoras quando discutem as ondas estacionárias no tubo, com exceção das coleções de Ramalho et al e Gaspar. A discussão realizada neste último é a menos cuidadosa. Além de não salientar a natureza longitudinal das ondas sonoras na seção onde as ondas sonoras estacionárias são discutidas, há ainda uma confusão à respeito das quantidades físicas em questão. São empregados os termos "rarefação" e "compressão" para a imagem que representa o deslocamento de ar. Entretanto, sem a separação entre deslocamento de ar e variação de pressão, estes termos podem confundir mais do que explicar.

Deve-se salientar que nenhum dos livros avaliados discute o comportamento oscilatório das quantidades medidas, nem a necessidade de se considerar ondas sonoras planas no interior do tubo.

\section{Sugestões}

Para tornar a discussão sobre o assunto mais clara, apresentamos algumas sugestões:

\begin{itemize}
\item Elaboração de experimentos;
\item O uso de aplicativos, applets, gifs, etc.;
\item Tópicos interessantes que podem ser discutidos.
\end{itemize}

Sugerimos a elaboração de dois experimentos, onde é possível evidenciar grandezas físicas importantes que não são devidamente discutidas nos livros didáticos (veja a seção \ref{ava_livro}). O primeiro trata da visualização do perfil  da pressão do ar e o segundo se destina à visualização do perfil do deslocamento do ar ao longo de um tubo com a uma extremidade aberta e outra fechada. 

Tendo isto em mente, elaboramos um arranjo experimental simples que permite observar o perfil de ondas sonoras estacionárias formadas em um tubo semi aberto. Uma grande vantagem desta montagem é que ela pode ser reproduzida não só em um laboratório, mas também em sala de aula e outros ambientes.


\subsection{Experimentos}
Para reproduzir os experimentos que discutimos neste trabalho o leitor deverá dispor da seguinte relação de materiais:

\begin{enumerate}[(i)]

\item Dois \textit{Tablets} (ou \textit{smartphones});
\item Um tubo de vidro aberto em uma extremidade e fechado na outra;
\item uma vareta de madeira (de tamanho compatível com o do tubo de vidro);
\item um alto-falante;
\item trena (ou régua);
\item fita adesiva;
\item um microfone;
\item pequenas bolas de isopor.

\end{enumerate}

Além desta estrutura física também faz-se necessário que os \textit{tablets} tenham alguns aplicativos previamente instalados.

\begin{figure}[!htb]
\begin{center}
\vspace{0.6cm}
\includegraphics[scale=0.25]{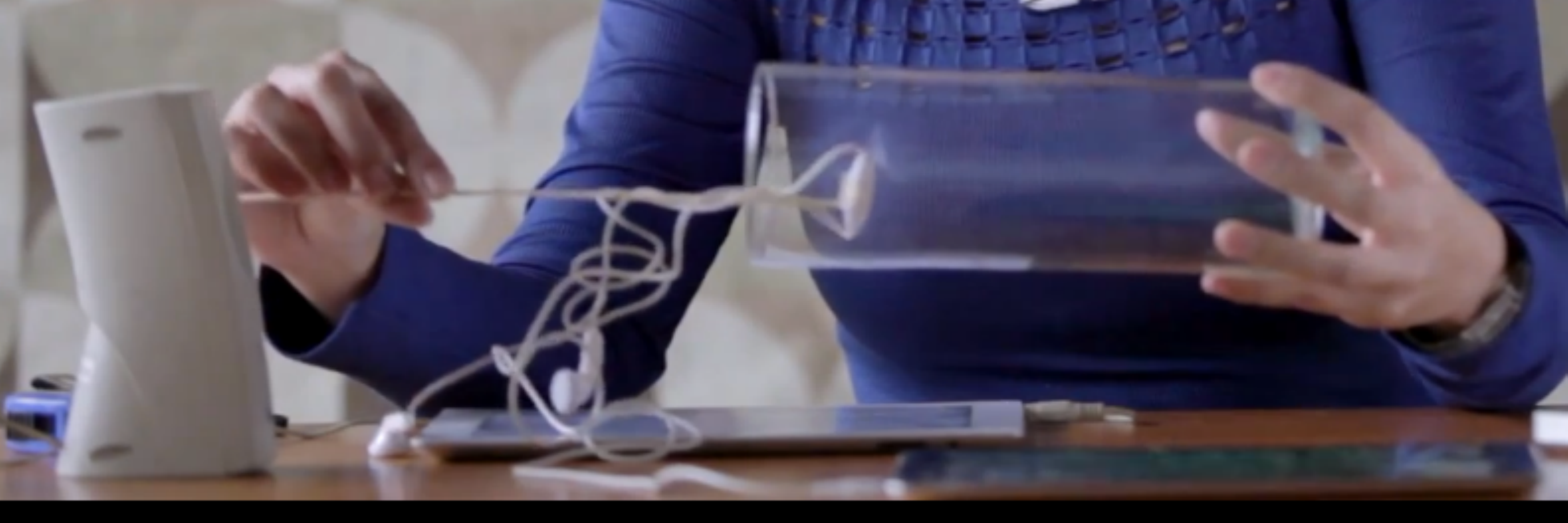}
\end{center}
\caption{Imagem que mostra a realização do experimento proposto neste trabalho.}
\label{montagem_exp}
\end{figure}

Para a reprodução do experimento utilizamos dois \textit{tablets}. Entretanto, nada impede que se utilizem dois \textit{smartphones}, ou um \textit{smartphone} e um \textit{tablet}. O importante é que os aplicativos necessários estejam instalados e funcionando devidamente [veja a Figura (\ref{montagem_exp}) onde mostramos o experimento proposto sendo realizado]. 

\begin{figure}[!htb]
\begin{center}
\vspace{0.6cm}
\includegraphics[scale=0.18]{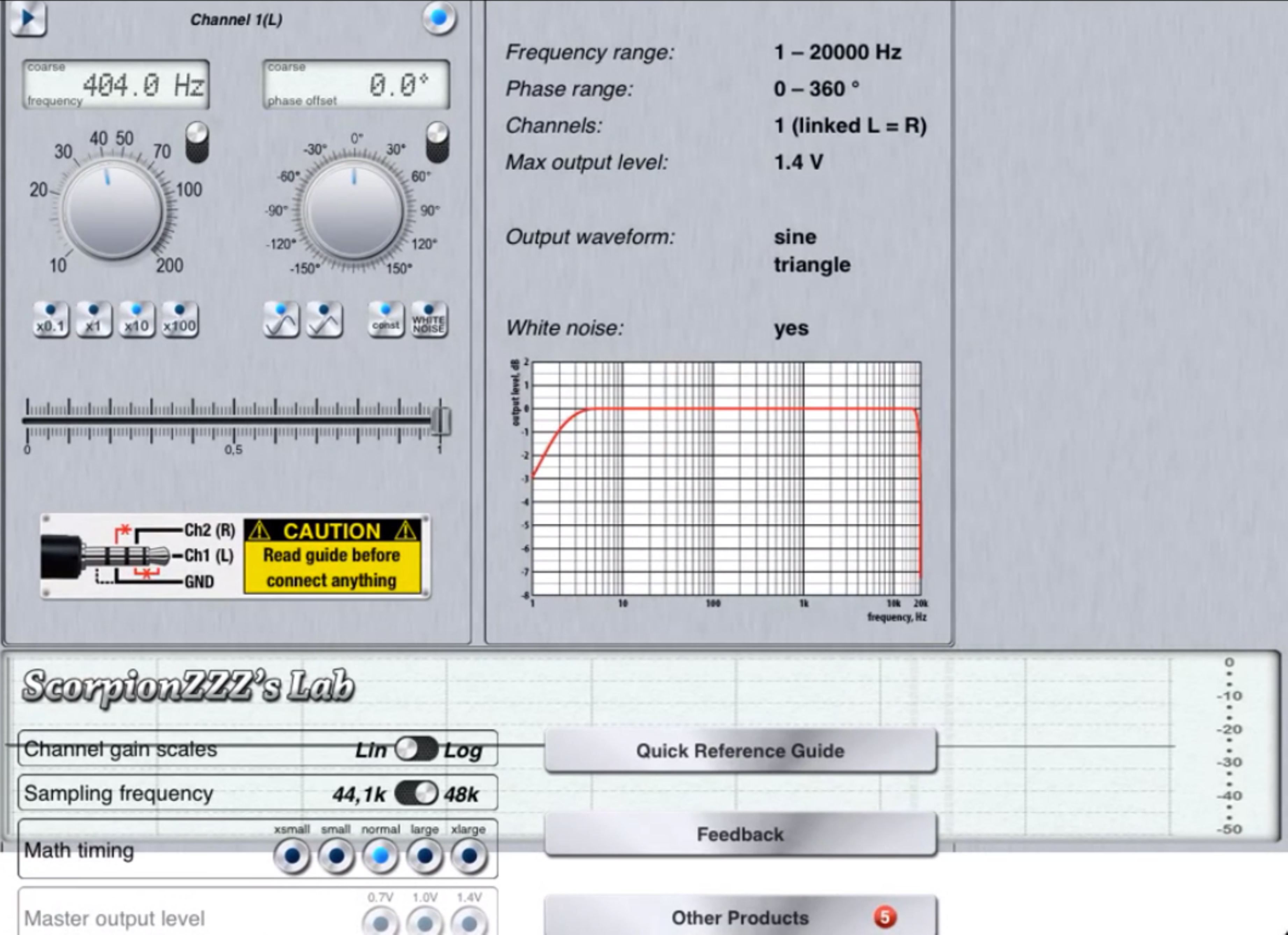}
\end{center}
\caption{Imagem capturada da tela do aplicativo \textit{SGenerator Lite}, a versão gratuita do \textit{SGenerator}.}
\label{sgenerator_screen}
\end{figure}

Um dos \textit{tablets} servirá como uma fonte sonora, em conjunto com o alto-falante. Neste caso utilizamos o aplicativo \textit{SGenerator Lite} \cite{sgenerator}, que embora seja gratuito e possua menos recursos que a versão paga, já nos permite escolher a intensidade e a frequência da onda sonora a ser gerada. Conectando o alto-falante ao \textit{tablet} já com o \textit{SGenerator Lite} aberto, você terá em mãos um gerador de sinal [veja a Figura (\ref{sgenerator_screen})].

Já o outro \textit{tablet} será responsável por realizar medidas da onda sonora formada no interior do tubo. Para isto instalamos no mesmo o aplicativo \textit{oScope Lite}\cite{oscope}, que é gratuito, e conectamos o microfone, que está preso à vareta por meio da fita adesiva. Uma vez que o aplicativo \textit{oScope Lite} está aberto, podemos passear com a vareta no interior do tubo de vidro e observar padrões de máximos e mínimos.



\begin{figure}[!htb]
\begin{center}
\vspace{0.6cm}
\includegraphics[scale=0.18]{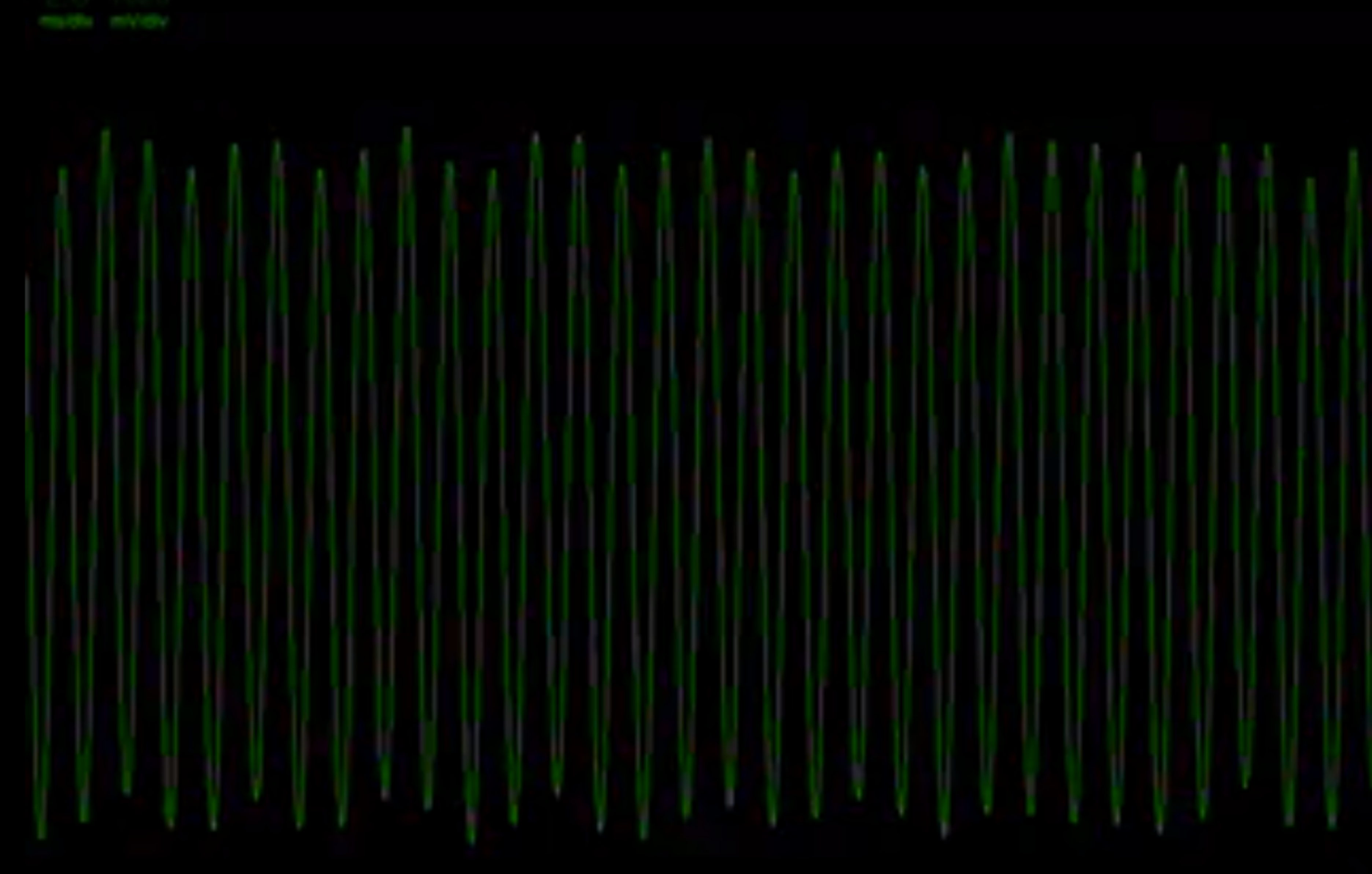}
\end{center}
\caption{Imagem capturada da tela do aplicativo \textit{oScope Lite}, versão gratuita do \textit{oScope}.}
\label{oscope_screen}
\end{figure}

Deste modo, é importante ressaltar que o perfil do harmônico que será visualizado pelo aplicativo \textit{oScope Lite} é o perfil da variação da pressão do ar com relacão ao eixo perpendicular da secção reta transversal do tubo, ou seja, uma excitação mecânica que é interpretada pelo aplicativo como a intensidade, ou grosso modo volume, da onda sonora [veja a figura (\ref{oscope_screen})]. Também seria possível reproduzir o mesmo experimento em um tubo aberto em ambas as extremidades, o que não foi feito aqui.

Antes de mais nada, podemos facilmente estimar com boa precisão quais são as frequências dos harmônicos formados em um tubo semiaberto. Para isto, basta medir o comprimento do tubo com o auxílio de uma trena ou régua, e tendo em vista que a velocidade do som no ar é de aproximadamente $340$ $m/s$ obtemos a frequência do harmônico fundamental através da relação 	$f_{0} = v_{som}/4L$. Para os demais harmônicos, basta utilizar a expressão \cite{harmonicos}

\begin{eqnarray}
f_{n} = \Big(\frac{n}{4L}\Big) v_{som} ,  \nonumber \\
 \nonumber \\
n= 1, 3, 5,... 
\label{harminicos}
\end{eqnarray}

Primeiramente ajustamos a frequência da onda sonora emitida pelo alto-falante para o valor de $f_{0} = 404$ Hz, estimado a partir do valor do comprimento do tubo. Em seguida colocamos o tubo cilindríco em frente ao auto falante. À medida que passeamos com o microfone acoplado à vareta pelo interior do tubo visualizamos no aplicativo \textit{oScope Lite} o perfil do harmônico fundamental, que apresenta um único mínimo de intesidade, localizado na extremidade aberta e um único máximo na extremidade fechada.

Para visualizar o deslocamento de ar podemos espaçar pequenas bolas de isopor pela extensão do tubo e apontar o alto-falante para a extremidade aberta do tubo. Para que possamos visualizar esta grandeza, é fundamental que o alto-falante tenha uma potência razoável em torno de 10 {\it W} e que as bolinhas estejam suficientemente espaçadas, de modo que a inércia das mesmas não atrapalhe a sua movimentação.

Para reforçar o caráter longitudinal da onda sonora no ar sugerimos a animação encontrada no site de D. Russel \cite{site_d_russell}.

\begin{figure}[!htb]
\begin{center}
\vspace{0.6cm}
\includegraphics[scale=0.112]{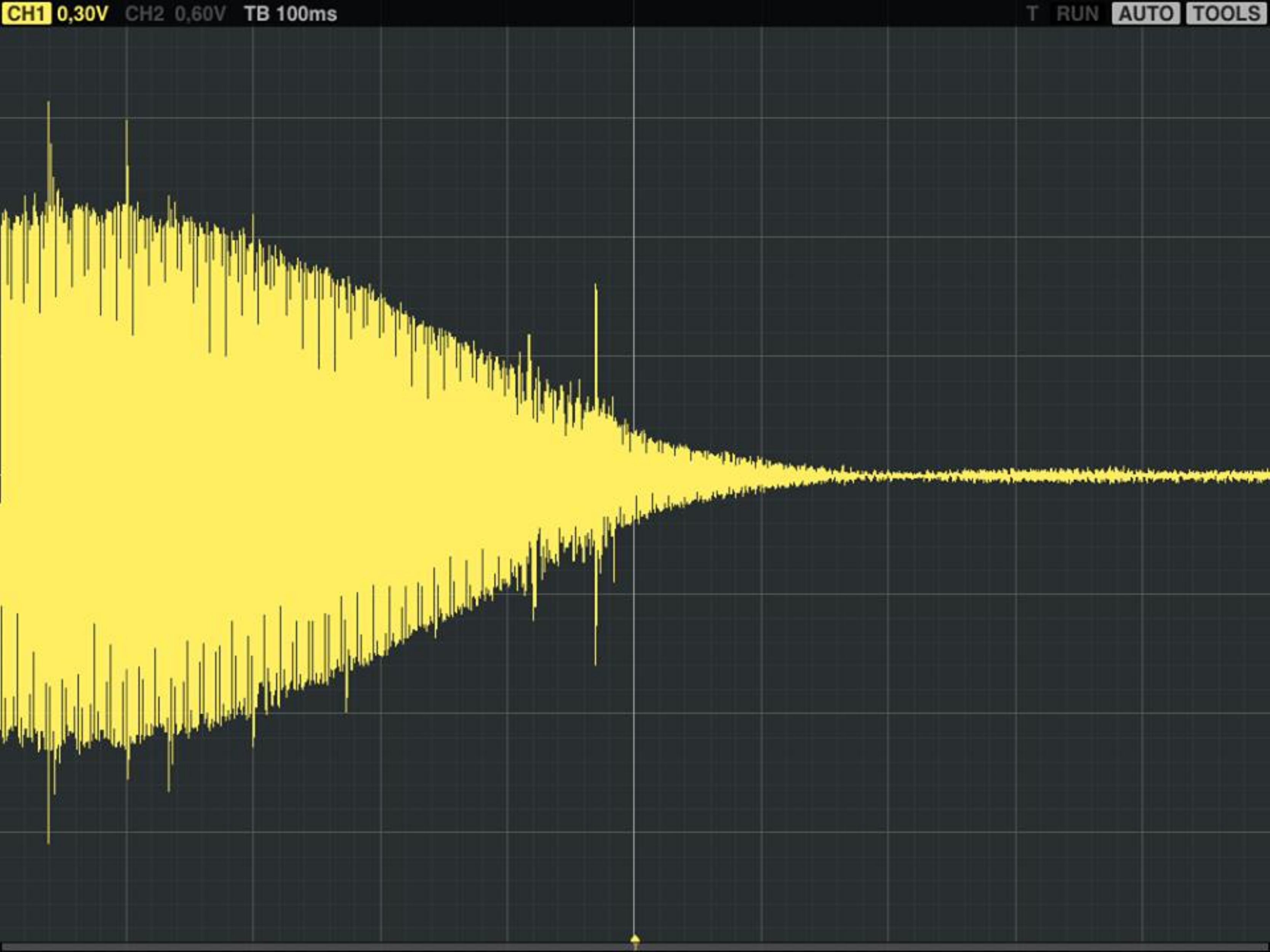}
\includegraphics[scale=0.112]{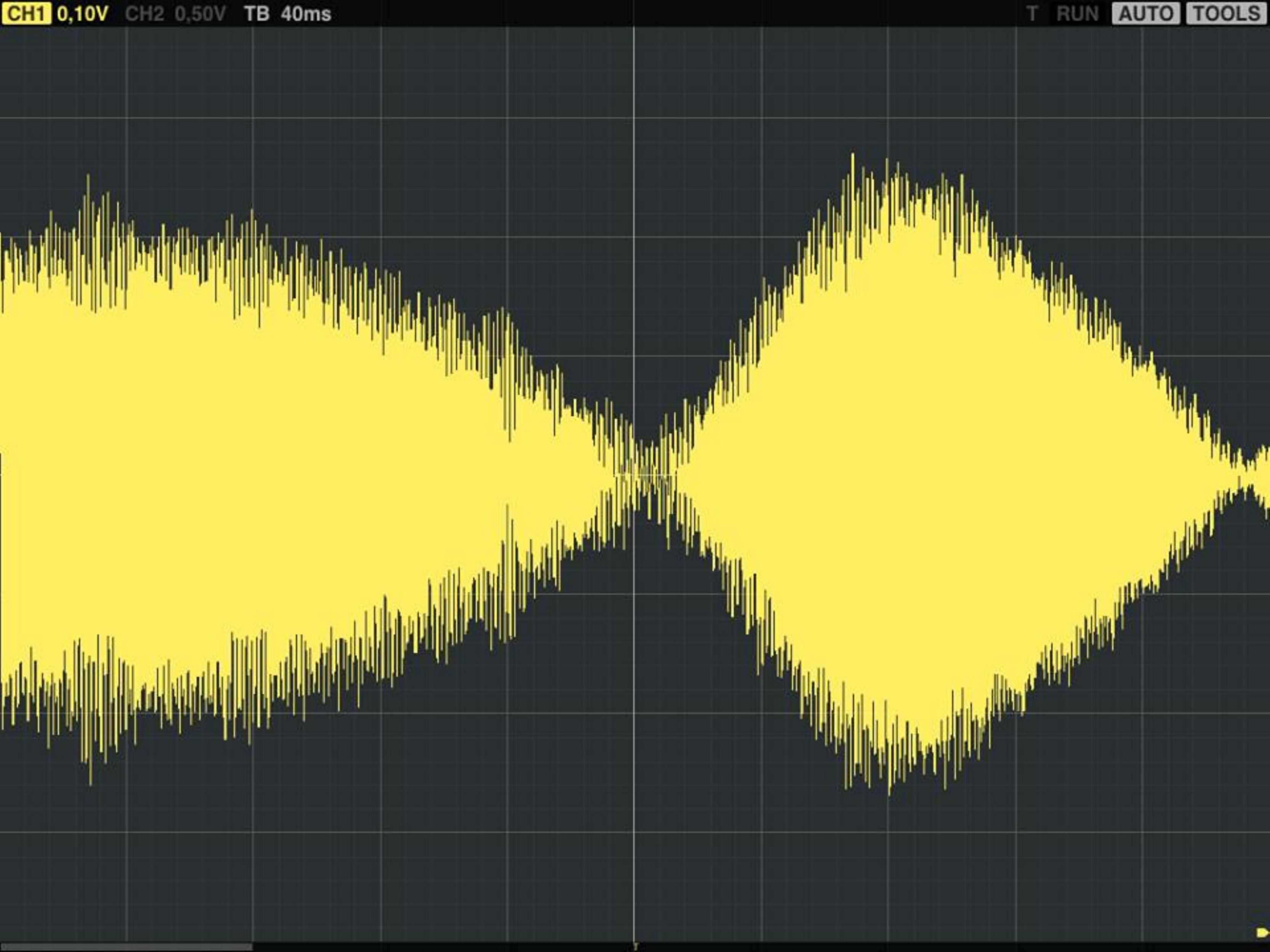}
\end{center}
\caption{Imagens obtidas com o aplicativo \textit{Oscilloscope} \cite{oscilloscope}. À esquerda temos o perfil da variação de pressão num tubo semi aberto para $f = 404 \, Hz$ ao longo do eixo do tubo. À direita, o mesmo para $f = 1212 \, Hz$.}
\label{figura_harmonicos}
\end{figure}

Reproduzimos também o segundo harmônico possível para o tubo semiaberto. Para isto, utilizamos a equação (\ref{harminicos}), obtendo $f_{1} = 1212$ Hz. Os padrões encontrados podem ser vistos na figura (\ref{figura_harmonicos}).

A reprodução de alguns dos demais harmônicos pode ficar um pouco comprometida a medida que as estimativas das frequências associadas à cada um deles necessitam de uma precisão cada vez maior. Nós conseguimos obter com cuidado suficiente ao menos os próximos dois harmônicos, além dos dois primeiros discutidos aqui.

%

\section{Conclusões e Perspectivas}

Neste trabalho apresentamos uma montagem experimental bastante simples e interessante, que permite ao professor e/ou estudante reproduzir ondas sonoras estacionárias em um tubo semiaberto. Além de propiciar a visualização de um fenômeno que normalmente desperta muitas dificuldades em uma abordagem tradicional, podemos discutir a natureza longitudinal de uma onda sonora no ar e apontar os problemas que normalmente são encontrados em livros texto de Ensino Médio e Superior. 

Pretendemos futuramente estender o experimento apresentado aqui para o caso de outras geometrias possíveis, tais como a formação de harmônicos sonoros em uma sala de aula ou corredor, por exemplo.

\section*{Saiba mais}

O leitor interessado pode assistir à uma série de vídeos produzidos pelos autores deste trabalho, onde apresentamos o experimento mostrado neste trabalho e muitos outros.

\url{http://www.youtube.com/channel/UC7E_sQiahyAzwd4FiUDhJxg}

\section*{Agradecimentos}

Os autores são gratos à Agência de Fomento CAPES e ao professor Anderson Ribeiro de Souza do colégio Pedro Segundo, Niterói, RJ.

\bibliographystyle{plain}

\begin{thebibliography}{40}

\bibitem{bybee} R. W. Bybee, G. E. Deboer, \textquotedblleft \textit{Research on goals for the science curriculum}\textquotedblright , \textit{Handbook of Research on Science Teaching and Learning},  p. 357-387, McMillan, 1996.

\bibitem{krasilchik} M. Krasilchik, \textquotedblleft Reformas e Realidade - o caso do ensino de Ciências. \textquotedblright , São Paulo em Perspectiva, v. 14, n 1, p. 85-93, 2000.

\bibitem{longhini} M. D. Longhini, \textquotedblleft O Uno e o Diverso na Educa\c{c}\~ao\textquotedblright , EDUFU, 2011.

\bibitem{borges} A. T. Borges, \textquotedblleft Novos rumos para o laboratório escolar de ciências\textquotedblright, Caderno Brasileiro de Ensino de Física, capa, v. 19, n. 3 (2002) 

\bibitem{previous_mag} N. Silva, \textquotedblleft \textit{Magnetic field sensor}\textquotedblright, \textit{The Physics Teacher}, v. 50, p.372-373 (2009). 

\bibitem{accelerometer} Página do \textit{MagnetMeter} na \textit{Apple Store}  \textit{\url{https://itunes.apple.com/us/app/magnetmeter-3d-vector-magnetometer/id346516607?mt=8}}. Acesso em 03/08/2013.


\bibitem{artigo_gota} L. P. Vieira, V. O. M. Lara, \textquotedblleft Macrofotografia com um \textit{tablet}: aplica\c{c}\~oes ao Ensino de Ci\^encias \textquotedblright , versão arxiv: \textit{\url{http://arxiv.org/abs/1307.4345}}.


\bibitem{site_d_russell} D. A. Russell, \textquotedblleft \textit{Acoustics and Vibration Animations}\textquotedblright \textit{\url{http://www.acs.psu.edu/drussell/Demos/StandingWaves/StandingWaves.html}}. Acesso em 03/08/2013.

\bibitem{maximo_alvarenga} A. Máximo, B. Alvarenga, \textquotedblleft F\'isica - Livro do Professor\textquotedblright, 1 ed., Vol. 2, Ed. Scipione , (2007).

\bibitem{guimaraes_boa} L. A. Guimarães, M. F. Boa, \textquotedblleft Física: Termologia, Óptica e Ondas\textquotedblright, 2 ed., Ed. Futura , (2004).

\bibitem{ramalho} F. Ramalho Jr., G. F. Nicolau, P. A. de Toledo, \textquotedblleft Os Fundamentos da Física\textquotedblright, 6 ed., v. 2, Ed. Moderna , (1997).

\bibitem{gaspar} A. Gaspar, \textquotedblleft Os Fundamentos da Física\textquotedblright, 1 ed., v. 2, Ed. Ática, (2000).

\bibitem{helou} R. Helou D., Gualter J. B., Newton V. B., \textquotedblleft Tópicos de Física\textquotedblright, ed. 18, v. 2, Ed. Saraiva , (2011).

\bibitem{hewitt} P. G. Hewitt, \textquotedblleft Física Conceitual\textquotedblright, ed. 11, Ed. Bookman , (2011).

\bibitem{gref} A. C. Copelli et al, \textquotedblleft Leituras de Física - GREF - Eletromagnetismo\textquotedblright, v. 3 Ed. EDUSP , (2005).

\bibitem{sgenerator} V. Korniienko, página para \textit{download} do \textit{SGenerator Lite} na \textit{Apple Store}  \textit{\url{https://itunes.apple.com/br/app/sgenerator-lite/id545708475?mt=8}}. Acesso em 03/08/2013.

\bibitem{oscope} A. Wiltschko, página para \textit{download} do \textit{oScope Lite} na \textit{Apple Store}  \textit{\url{https://itunes.apple.com/br/app/oscope-lite/id373858824?mt=8}}. Acesso em 03/08/2013.

\bibitem{harmonicos}  D. Halliday; R. Resnick e K. S. Krane , \textquotedblleft Física 2\textquotedblright , ed. 7, v. 2, LTC \& Sons (1996).

\bibitem{oscilloscope} Página do \textit{Oscilloscope} na \textit{Apple Store}  \textit{\url{https://itunes.apple.com/br/app/oscilloscope/id388636804?mt=8}}. Acesso em 03/08/2013.

\end{thebibliography}


\end{document}